\documentclass[aps,pra,twocolumn,noshowpacs,superscriptaddress]{revtex4-1}
\usepackage{graphicx,epsfig, subfigure}
\usepackage{amsbsy}
\usepackage{amssymb}
\usepackage{amsmath}
\usepackage{hyperref}
\usepackage[bottom]{footmisc}
\usepackage{color}
\usepackage{verbatim}

\date{\today}

\begin{document}
\title{Thermal Atom-Ion Collisions in K-Yb$^{+}$ Hybrid System}

\author{Thai M. Hoang}

\author{Peter D. D. Schwindt}

\author{Yuan-Yu Jau}
\affiliation{Sandia National Laboratories, Albuquerque, NM 87123, USA}

\begin{abstract}
We present experimental studies of atom-ion collisions using buffer-gas cooled, trapped ytterbium (Yb$^+$) ions immersed in potassium (K) vapor. The range of the collisional temperature is on the order of several hundred kelvin (thermal regime). We have determined various collisional rate coefficients of the Yb$^+$ ion per K-atom number density. We find the upper bounds of charge-exchange rate coefficients $\kappa_{\rm ce}$ to be $(12.7\pm1.6)\times10^{-14}$ cm$^3$s$^{-1}$ for K-$^{171}$Yb$^+$ and $(5.3\pm0.7)\times10^{-14}$ cm$^3$s$^{-1}$ for K-$^{172}$Yb$^+$. For both isotopes, the spin-destruction rate coefficient $\kappa_{\rm sd}$ has an upper bound at $(1.46\pm0.77)\times10^{-9}$ cm$^3$s$^{-1}$. The spin-exchange rate coefficient $\kappa_{\rm se}$ is measured to be $(1.64\pm0.51)\times10^{-9}$ cm$^3$s$^{-1}$. The relatively low charge-exchange rate reported here demonstrates the advantage of using K atoms to sympathetically cool Yb$^+$ ions, and the relatively high spin-exchange rate may benefit research work in quantum metrology and quantum information processing on an atom-ion platform using K atoms and Yb$^+$ ions.
\end{abstract}

% make the title area
\maketitle

\section{Introduction}
While free thermal atoms, trapped neutral atoms, and trapped ions are well established platforms for atomic-physics research and applications, a hybrid system of neutral atoms and ions can offer new capabilities and opportunities that cannot be accomplished with only one atomic species. In 1960s and 1970s, the spin-dependent charge exchange between $^3\mathrm{He}^+$ ions and $\mathrm{Cs}$ atoms was used to measure the ground-state hyperfine splitting frequency of $^3\mathrm{He}^+$ ions, because the narrow-band light source at 41-eV photon energy was not available to directly probe the internal state of $^3\mathrm{He}^+$ \cite{fortson1966ultrahigh, schuessler1969hyperfine}. Up to date, this Cs mediated method still delivers the best result of the $^3$He$^+$ ground-state hyperfine frequency. For similar reasons, the spin-exchange interactions between free ions and $\mathrm{Rb}$ atoms were used to perform hyperfine spectroscopy of $\mathrm{Sr}^+$,  $\mathrm{Cd}^+$, and  $\mathrm{Hg}^+$ ions \cite{gibbs1971production, schuessler1973measurement} due to the unavailable blue to ultra violet (UV), narrow-line light sources for optical pumping. These atom-ion hybrid systems demonstrated a great potential for atomic/quantum metrology applications such as fieldable ion clocks \cite{Jau12}, eliminating the need of miniaturized, low-power, short-wavelength ($\lesssim400$ nm) laser sources for interrogating ion systems. In the past two decades, a hybrid system of cold neutral atoms and cold ions has become an attractive platform to study quantum physics of heterosystems \cite{tomza2018, Furst2018-2}, which will advance our knowledge in cold collisions \cite{Ravi2012}, quantum chemistry, many-body physics, quantum simulation \cite{tomza2018}, and quantum information processing \cite{doerk2010atom}.

In the past decades, various collisional phenomena in atom-ion systems have been investigated broadly through theoretical modeling (see citations in Ref\cite{tomza2018}) and experiments\cite{Smith2005, grier2009observation, zipkes2010cold, zipkes2010trapped, schmid2010dynamics, Rellergert2011, Ravi2012, Ratschbacher2013, Smith2014, Goodman2015, Furst2018, Sikorsky2018}. Still, there are many atom-ion pairs that have not been studied, and more relevant experimental research work will be necessary. In this paper, we study the K-Yb$^{+}$ hybrid system using helium buffer-gas cooled, trapped ytterbium (Yb$^+$) ions and free potassium (K) atoms (vapor). Potassium offers one of the few fermion atomic species and therefore is interesting to study. To our knowledge, there are no theoretical and experimental results of the K-Yb$^{+}$ system being reported previously. Through our experiments, we determine the charge-exchange, spin-destruction, and spin-exchange rate coefficients for $\mathrm{Yb}^+$ ions and $\mathrm{K}$ atoms. Ideally, if the collisions follow the Langevin model\cite{tomza2018}, the rate coefficients are nearly independent of the collisional kinetic energy. While the previous experimental studies of hybrid atom-ion systems used both cold ions and cold atoms \cite{Smith2005,grier2009observation,zipkes2010cold,zipkes2010trapped,schmid2010dynamics,Ratschbacher2013,Smith2014, Sikorsky2018}, or thermal ions ($>100$ kelvin) \cite{Goodman2015} and cold atoms, our experiment was carried out with both thermal ions and thermal atoms. This would help with further understanding of the collisional interactions of hybrid systems in a wider temperature range.

\begin{figure*}[t]
%\centering
\includegraphics[scale=0.35]{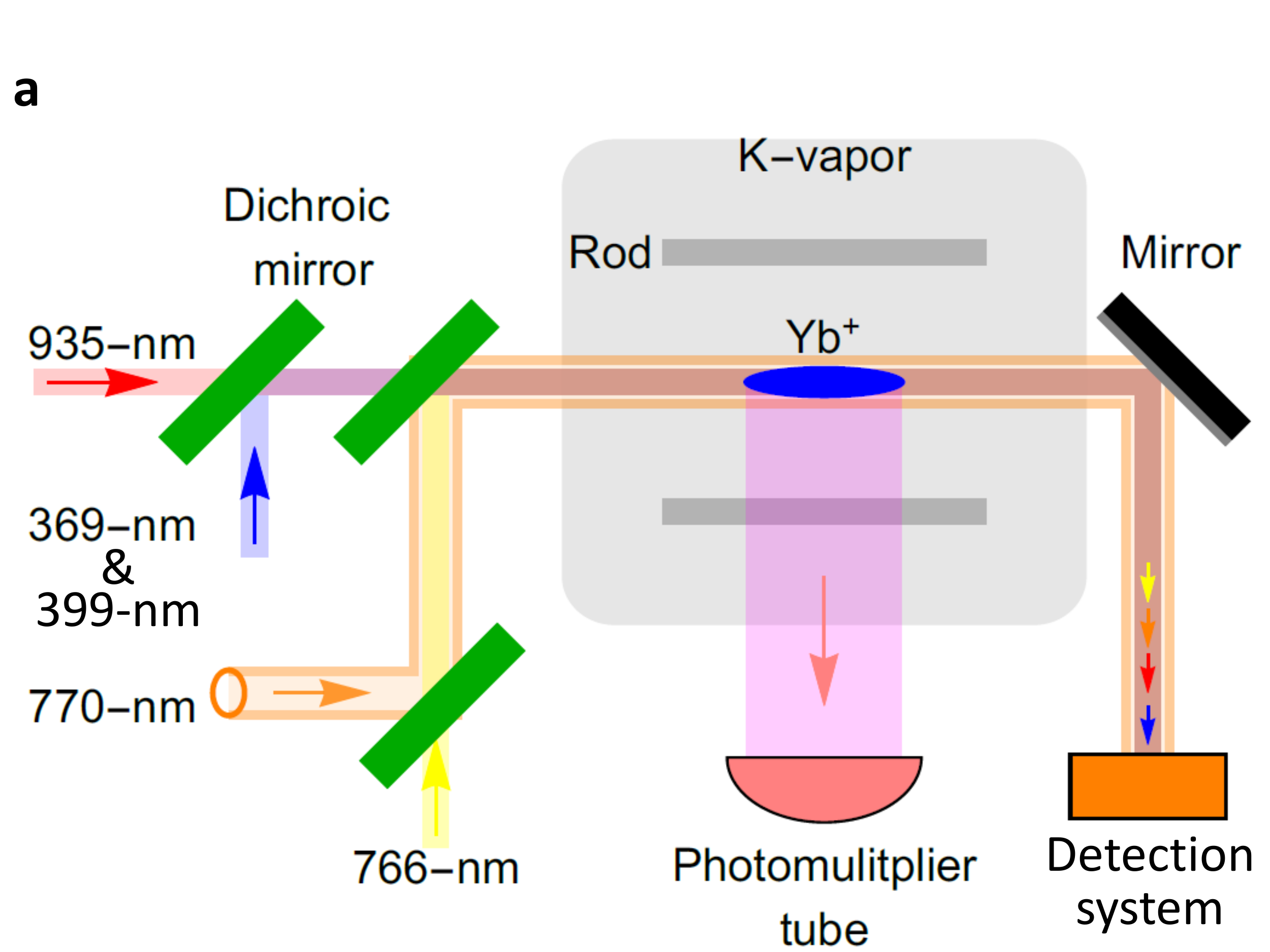}
\includegraphics[scale=0.7]{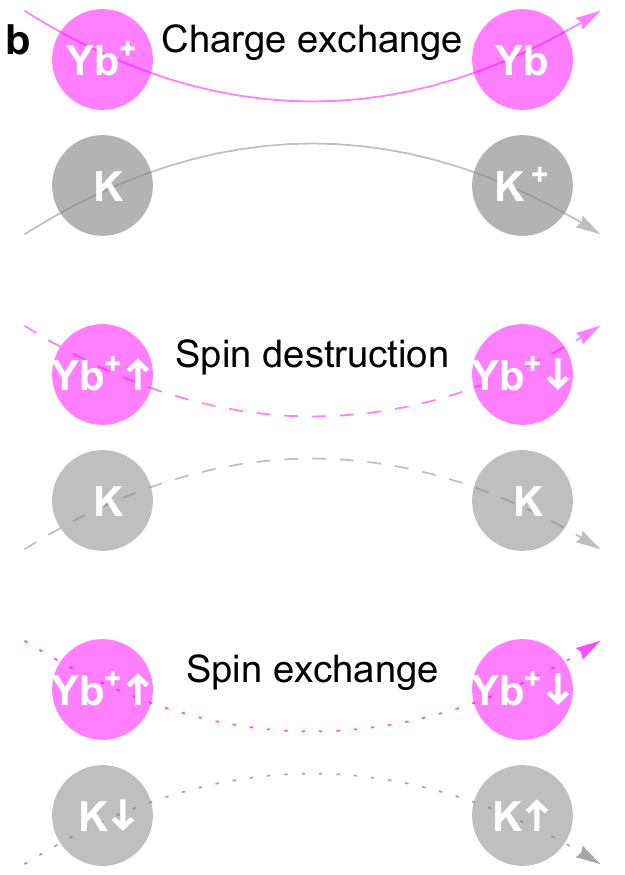}
\includegraphics[scale=0.7]{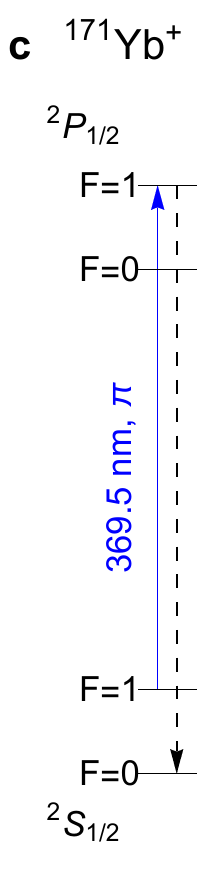}
\includegraphics[scale=0.7]{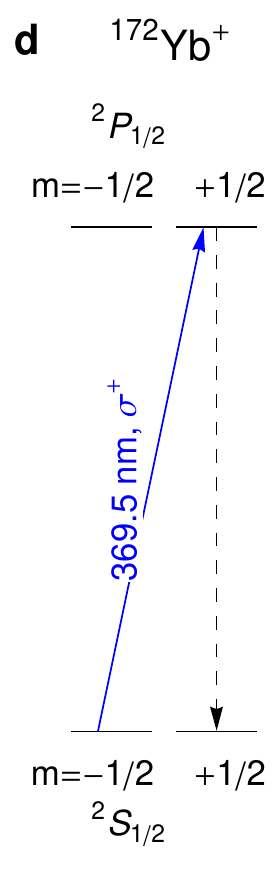}
% where an .eps filename suffix will be assumed under latex,
% and a .pdf suffix will be assumed for pdflatex; or what has been declared
% via \DeclareGraphicsExtensions.
\caption{a) The Yb$^+$ ions cloud (blue) are trapped in a linear Paul trap. The 369-nm, 399-nm, 935-nm, 770-nm, and 766-nm laser beams are combined using dichroic mirrors. The fluorescence signal of the Yb$^+$ ions is collected by a PMT. The laser beams go to a detection system with different wavelength filters. K atoms (gray area) are released into the chamber from a potassium getter. b) Illustrations of the charge-exchange, the spin-destruction, and the spin-exchange interactions. The up and down arrows represent an atomic spin. c) Hyperfine polarization of $\mathrm{^{171}Yb^+}$ ions achieved by optical pumping. d) Spin polarization of $\mathrm{^{172}Yb^+}$ ions achieved by optical pumping.}
\label{Apparatus}
\end{figure*}

\section{Apparatus and Settings}
Our experimental apparatus comprises a vacuum chamber for the atom-ion hybrid environment, B-field coils, and various laser sources. Inside the vacuum chamber, there is a linear Paul trap with four 50-mm long rods (6-mm apart along each side) and two hollow end caps, and there are two Yb sources (natural abundance and isotropically enriched) and one natural-abundance K-getter source (i.e. a potassium dispener). The chamber was baked out to achieve $10^{-10}$ torr level and then was back filled with $45\times 10^{-6}$ torr of helium gas for buffer-gas cooling to the trapped Yb$^{+}$ ions. The K getter was driven by 2.5-A DC current during the bakeout to minimize the possible out-gassing sources on the getter. The trap rods were driven by a 1.95-MHz RF source at  $780~\mathrm{V_{pk-pk}}$ and the end caps were connected to a 15-V DC source. We were able to trap $>10^6$ ions with an ion density on the order of $10^8$ cm$^{-3}$.  The Yb$^+$ ion temperature was measured to be around 650 K. When the trap was loaded with ions, the radial Secular frequency was around 200 kHz, which is consistent with the simulation results from our 3D ion-trap modeling including space-charge effect.

We used laser sources with five different wavelengths for experimental studies. The laser sources at 369 nm, 399 nm, and 935 nm are for ion loading, state preparation, and interrogation of Yb$^{+}$ ions\cite{Jau12,Jau15}. The laser sources at 766 nm and 770 nm are for optical pumping and probing of K atoms. As illustrated in Fig.~\ref{Apparatus}a, a 369-nm laser beam is delivered to the ion cloud to provide state detection and optical pumping of the trapped ions, and the 935-nm laser clears the low-lying $D_{3/2}$ state. Fluorescence at 369 nm or 297 nm is collected with a photomultiplier tube (PMT). For loading the Yb$^+$ ions, the combination of the 399-nm laser and 369-nm laser allows us to selectively load $^{171}$Yb$^+$ ions or $^{172}$Yb$^+$ ions into the ion trap. To study the interactions between $\mathrm{Yb}^+$ ion and K atoms, the vacuum chamber was filled with K vapor released from a potassium getter. The K vapor density was controlled through the electrical current applied on the getter. The measured K-atom temperature was around 340 K. In Fig.~\ref{Apparatus}a, a linearly polarized 770-nm laser beam with a doughnut profile is used as a probe to measure the K-vapor density and temperature. The same laser beam with circular polarization is also used to optically spin-polarize K atoms. The dark center of the doughnut beam is aligned to the ion cloud axis. A linearly polarized 766-nm laser beam through the ion cloud is used for an optical rotation (Faraday effect) measurement\cite{Mathur1970,Happer2010} to determine the degree of spin polarization of the K atoms inside the ion cloud. The number density and the temperature of the trapped Yb$^+$ ions or the free-space K atoms can be determined by measuring the signal contrast and the Doppler profile of the $D1$ resonances through scanning the frequency of the 369-nm or the 770-nm probe laser \cite{Mathur1970,Happer2010}, and the measurement uncertainty was within a few percent. In the detection system, we used single photodetector to detect the transmission light at 369 nm and 770 nm, and we used a two-photodiode balanced detector with an optical polarimetry setup to measure the Faraday rotation signals from K atoms at 766 nm. We used dichroic mirrors, spectral filters, and flipping mirrors to steer and manage different laser beams in the detection system. We used three orthogonal coil pairs (Helmholtz-like) to compensate the ambient magnetic field and to set the B-field magnitude to be around 1 G with preferred orientations for measurements.

\section{Measurements and Results}
\subsection{Charge exchange}
Figure~\ref{Apparatus}b illustrates different K-Yb$^{+}$ collisional interactions, such as charge exchange, spin destruction, and spin exchange.
The charge-exchange interaction occurs when an electron migrates from a neutral K atom to a Yb$^+$ ion during the collision with or without an involvement of emitting a photon (There could be an intermediate state $(\rm YbK)^+$ during the collisional process.):
\begin{eqnarray}
	\text{charge exchange: }\mathrm{Yb}^+ + \mathrm{K} \rightarrow 	\mathrm{Yb} + \mathrm{K}^+. \nonumber
\end{eqnarray}
After charge-exchange collisions, the neutral Yb atom can no longer be trapped. This is the main ion loss mechanism. Ideally, we can measure the charge-exchange-induced ion-loss rate depending on the K-atom number density to determine the charge-exchange rate coefficient via the following relation:
\begin{eqnarray}
	\gamma_{\rm ce}=\frac{1}{\tau_{\rm ce}}=\kappa_{\rm ce}\cdot n_\mathrm{K},
	\label{Eqn:ChargeExchange}
\end{eqnarray}
where $\gamma_{\rm ce}$ is the charge-exchange rate, $\tau_{\rm ce}$ is the exponential ion loss time constant, $n_\mathrm{K}$ is the number density of K atoms, and $\kappa_{\rm ce}$ is the charge-exchange rate coefficient. Here, the rate coefficient $\kappa=\langle\sigma v\rangle$ is an ensemble average of all possible velocity-dependent cross section $\sigma(v)$ and colliding relative velocity $v$.

To measure $\tau_{\rm ce}$, we delivered a 1-mm diameter ($1/e^2$), 3-mW continuous 935-nm laser beam overlapped with a 1-mm diameter ($1/e^2$) pulsed 369-nm laser beam to the ion cloud. The 369-nm beam was linearly polarized to avoid spin-polarized optical pumping and was pulsed every 11 minutes. We verified that the 369-nm laser power and its duty cycle were sufficiently low to eliminate any possible 369-nm laser associated ion-signal-loss effects, such as F-state trapping \cite{Jau15}, laser induced molecular ion formation from Yb$^+$ ions and the background gas \cite{Sugiyama1997}, and laser-enhanced charge exchange (explained in the end of this section). For $^{171}\mathrm{Yb}^+$ ions, we used $F=1$ to $F'=0$ transition to excite the ions and generate fluorescence. By taking advantage of this cycling transition \cite{Jau12}, we can achieve extremely slow hyperfine optical pumping. We found that for the 369-nm laser power $\leq1$ $\mu$W and the pulse duration $\leq4$ seconds, the measured ion lifetime has no noticeable dependence on the laser power and pulse duration. For data analysis, we include experimental data with laser power at 0.2 $\mu$W, 0.5 $\mu$W, and 1 $\mu$W and with pulsewidth at 0.2 s, 0.4 s, 1 s, 2 s, and 4 s.

\begin{figure}[t]
%\centering
\includegraphics[scale=0.85]{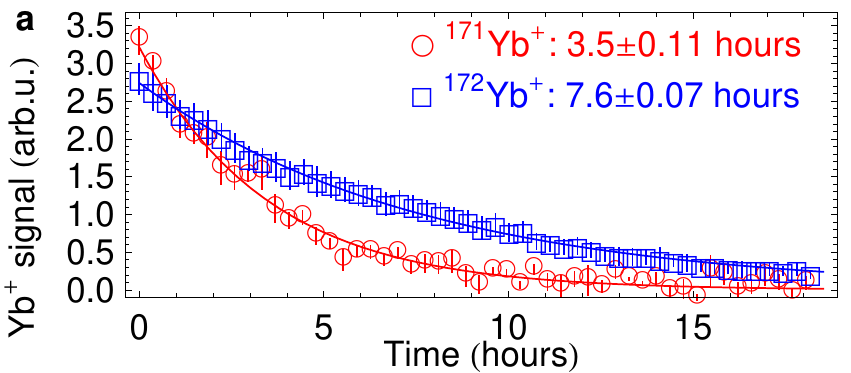}
\includegraphics[scale=0.85]{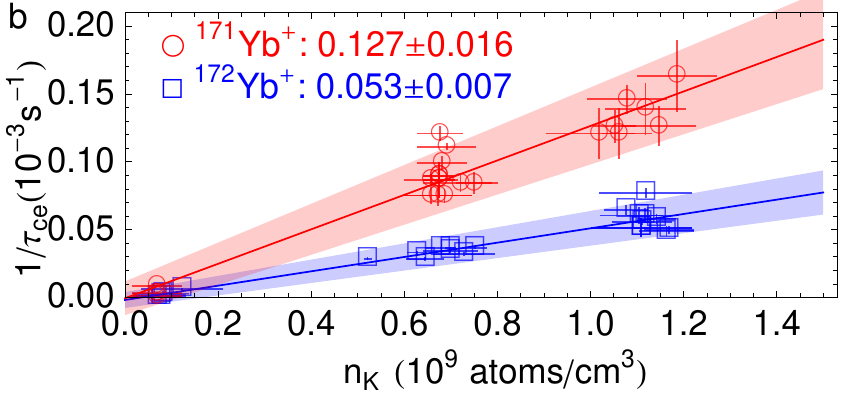}
% where an .eps filename suffix will be assumed under latex,
% and a .pdf suffix will be assumed for pdflatex; or what has been declared
% via \DeclareGraphicsExtensions.
\caption{a) Decay signals of Yb$^+$ ions due to ion loss under the presence of $\sim7\times10^8$ cm$^{-3}$ K-atom number density. The data are fitted to an exponential function to extract $\tau_{\rm ce}$. The numbers represent the charge-exchange time constants and the fitting errors. b) The charge-exchange rate of a Yb$^+$ ion under different potassium densities. Experimental data of $^{171}\mathrm{Yb}^+$ ($^{172}\mathrm{Yb}^+$) ions are shown in red circles (blue squares). The data are fitted to a linear function to extract the slope and to determine the upper bound of $\kappa_{\rm ce}$. The shaded areas represent 95\% confidence interval.}
\label{Charge}
\end{figure}

The typical, decaying fluorescence signals of Yb$^+$ ions are shown in Fig.~ \ref{Charge}a. To extract $\tau_{\rm ce}$, we fit the data to an exponential function. The data indicates that the $\mathrm{^{171}Yb^+}$ ions leave the trap about twice faster than $\mathrm{^{172}Yb^+}$ ions with the same K-atom density. The reason why is unclear at this point. Although $^{171}$Yb$^+$ ion has hyperfine structure in contrast to $^{172}$Yb$^+$ ion, the collisional energy in thermal regime is too large to differentiate the hyperfine states. In addition, both Yb$^+$ ions and K atoms mainly remain unpolarized throughout the entire period of measuring the ion loss rate, therefore it is unlikely to see polarization-dependent or state-dependent charge exchange effects, unless different hyperfine manifolds have very distinct charge-exchange cross sections. On the other hand, when checking the ion-trapping lifetime without the presence of K atoms (the K-getter is off) every once in a while over several months, the $\mathrm{^{172}Yb^+}$ ions demonstrated consistent trapping lifetime ($>100$ hours), but the trapping lifetime of the $\mathrm{^{171}Yb^+}$ ions can be sometimes longer and sometimes shorter with a difference up to a factor of 2. We suspect that the some kind of non-detectable molecular gas background in the chamber was more reactive with $\mathrm{^{171}Yb^+}$ ions and caused the phenomenon.  

Figure~\ref{Charge}b shows the summary of experimental data of $1/\tau_{\rm ce}$ for K-$^{171}$Yb$^+$ and K-$^{172}$Yb$^+$ with different K-atom density $n_{\rm K}$.  However the measured data of $1/\tau_{\rm ce}$ may not be completely dominated by $n_{\rm K}$. Although we had tried to clean up the getter during the bakeout, and there was no detectable signals from the gas residual analyzer (RGA) connected to the vacuum system, we still cannot rule out the possibility of introducing some gas background when running the K-getter, because the RGA was much farther from the ion trap and the K-getter was closer to the trap. This possible additional gas background can be positively correlated with $n_{\rm K}$ (determined by the getter temperature) and can react with Yb$^+$ ions to reduce the ion-trapping lifetime. Considering this systematic effect and the possible residual optical effect, with linear fits, we find the upper bounds of the charge-exchange rate coefficients to be $^{171}\kappa_{\rm ce}\leq(12.7\pm1.6)\times10^{-14}\text{ cm}^3\rm s^{-1}$ and $^{172}\kappa_{\rm ce}\leq(5.3\pm0.7)\times10^{-14}\text{ cm}^3\rm s^{-1}$. In Fig~\ref{Charge}b, the vertical error bars come from the exponential fitting errors as an example in Fig~\ref{Charge}a. The horizontal error bars come from the K-density difference before and after the each ion lifetime measurement was finished. The K-density measurement by itself has an uncertainty within a few percent, but the K-density can drift more than 10\% over an ion lifetime measurement.

It is worth noting that the ion loss rate under the presence of K atoms increases rapidly when either the Yb$^+$ ions or the K atoms were exposed to a strong 369-nm laser beam or a strong 770-nm laser beam at corresponding $D1$ transition wavelengths. We believe this is caused by a much stronger charge-exchange interaction when the colliding ion or atom is in the excited state. In our preliminary study, we checked the ion-trapping lifetime using a continuous 100 mW/cm$^2$ 369-nm laser beam (Gaussian profile) or a continuous 200 mW/cm$^2$ 770-nm laser beam (Gaussian profile) to illuminate the ion cloud, and the ion-trapping lifetime is significantly shorter (on the order of 10-100 seconds) when the K atoms ($\sim10^9$ cm$^{-3}$) are present. We estimated the charge-exchange rate coefficient to be on the order of $10^{-9}$ to $10^{-8}$ cm$^3$s$^{-1}$ if one of the colliding species is in its excited state $(P_{1/2})$. In the experiments of determining the ground-state charge-exchange rate coefficients, the highest 369-nm laser power was 1 $\mu$W, which leads to a relative population of $\sim10^{-5}$ in the $P_{1/2}$ state of Yb$^+$ ion (Note: due to the Doppler broadening of the thermal ions, the $D1$ on-resonance absorption cross section is a few times $10^{-12}$ cm$^2$). For the data shown in Fig.~\ref{Charge}b, the duty cycle of the 369-nm laser pulse is $\leq0.6$\%. The equivalent charge-exchange rate coefficient contributed from the $P_{1/2}$ state is therefore $\leq6\times10^{-(16\text{ to }17)}$  cm$^3$s$^{-1}$, which is negligible compared to the measured values.

\begin{figure}[t]
\centering
	\includegraphics[scale=0.85]{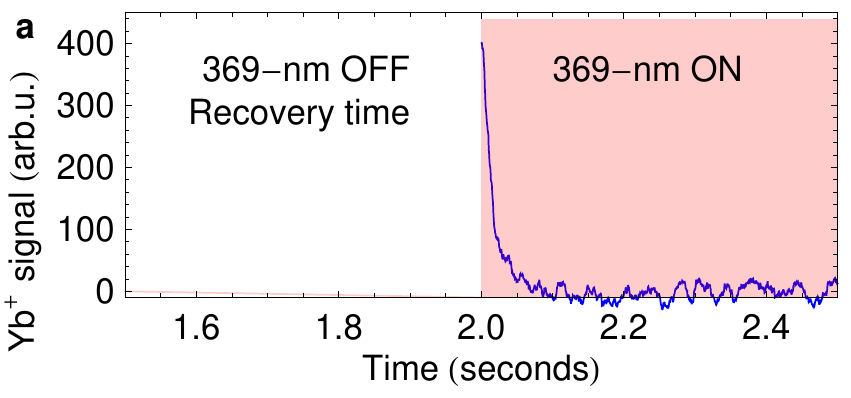}
	\includegraphics[scale=0.85]{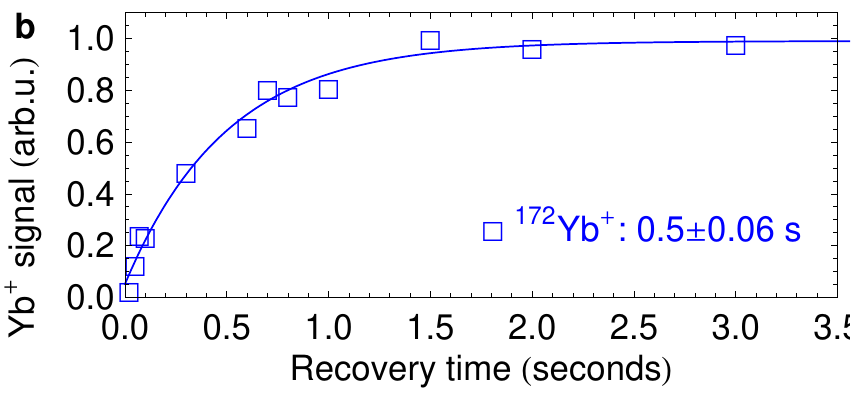}	
	\includegraphics[scale=0.85]{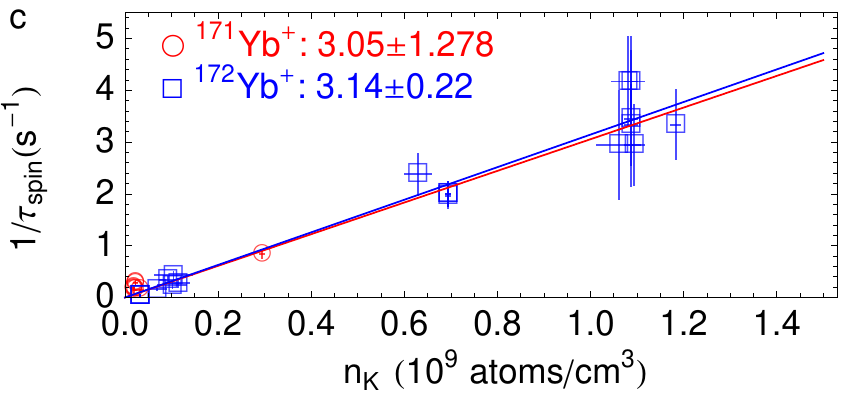}	
%\includegraphics[width=3.4in]{PowerPointSummary/Plot/SpinExchangePlot171Paper.pdf}
% where an .eps filename suffix will be assumed under latex,
% and a .pdf suffix will be assumed for pdflatex; or what has been declared
% via \DeclareGraphicsExtensions.
\caption{a) The red shaded area represents the 50-nW, 0.5-s pulse of the circularly polarized 369-nm laser light. b) The data points with a normalized scale represent the areas under of the fluorescence transient curves for different recovery time periods, where $n_K\sim7\times10^8$ cm$^{-3}$. The numbers represent the spin recovery time constants and the fitting errors. Each data point is an average of 200 runs. c) The spin recovery time constants versus different K number densities. Data of $^{172}\mathrm{Yb}^+$ ions  and $^{171}\mathrm{Yb}^+$ ions are shown in blue square markers  and red circle markers, respectively. The $^{171}$Yb$^+$ data uses similar procedure for  $^{172}$Yb$^+$ except that the 369-nm laser is linearly polarized at 5 $\mu$W with 3.5-s pulse width. The solid lines represent the linear fitting results. The fitting lines are with 95\% confidence level.}
\label{Spin}
\end{figure}

\subsection{Spin destruction and spin exchange}
Atom-ion collisions can induce not only charge exchange but also spin flips, including random spin flips that destroy the spin polarization of the colliding atomic species, and spin exchange that swaps the electronic angular momentum between the ion and the atom. These spin-dependent collisional effects on the Yb$^+$ ion leads to these forms:
\begin{eqnarray}
	\text{spin exchange: }&&\mathrm{Yb}^+\uparrow + \mathrm{K}\downarrow \rightarrow  \mathrm{Yb}^+\downarrow + \mathrm{K} \uparrow, \nonumber\\
	\text{spin destruction: }&&\mathrm{Yb}^+\uparrow\downarrow + \mathrm{K} \rightarrow 	 \mathrm{Yb}^+\downarrow\uparrow + \mathrm{K}. \nonumber
\end{eqnarray}
Here the $\uparrow, \downarrow$ arrows represent the spin state. The dynamics of the Yb$^+$ spin polarization affected by the 369-nm optical pumping and the K-atom density is described by
\begin{equation}\label{SpinDynamics}
%\dot{P}_{\rm Yb^+}=n_{\rm K}\left[-\kappa_{\rm sd}P_{\rm Yb^+}+\kappa_{\rm se}(P_{\rm K}-P_{\rm Yb^+})\right],
\dot{P}_{\rm Yb^+}=R_{\rm op}(1-P_{\rm Yb^+})+n_{\rm K}\left[\kappa_{\rm se}(P_{\rm K}-P_{\rm Yb^+})-\kappa_{\rm sd}P_{\rm Yb^+}\right],
\end{equation}
where $R_{\rm op}$ is the optical pumping rate on the ions, $P_{\rm Yb^+}$ and $P_{\rm K}$, ranging from -1 to 1, are the spin polarization for the Yb$^+$ ions and K atoms, and $\kappa_{\rm sd}$ and $\kappa_{\rm se}$ are the rate coefficients for the spin-destruction and spin-exchange interactions. From Eq.~\ref{SpinDynamics}, with $R_{\rm op}=0$ and $P_{\rm K}=0$, we find the total spin-relaxation time constant to be $\tau_{\rm spin}=\gamma_{\rm spin}^{-1}=[(\kappa_{\rm sd}+\kappa_{\rm se})n_{\rm K}]^{-1}$. To measure the total spin-relaxation rate $\gamma_{\rm spin}$, we keep the K atoms unpolarized (i.e. $P_{\rm K}=0$, no optical pumping on K atoms), and we spin-polarize the $^{172}$Yb$^+$ ions by applying a circularly polarized 369-nm laser to pump the $^{172}$Yb$^+$ ion into the $|m=1/2\rangle=|\uparrow\rangle$ or $|m=-1/2\rangle=|\downarrow\rangle$ state in its Zeeman sublevels as illustrated in Fig.~\ref{Apparatus}d. As shown in Fig.~\ref{Spin}a, when the unpolarized $^{172}$Yb$^+$ ions are illuminated by the 369-nm laser light, fluorescence appears at the beginning and then decays as most of the ions are spin polarized ($P_{\rm Yb^+}\sim1$), where ions are in the dark state of the pump light and the photon scattering is minimized. During the ``OFF'' period of the pump light (i.e. $R_{\rm op}=0$), the spin relaxation process reduces $P_{\rm Yb^+}$ and therefore recovers the fluorescence signal when the pump light turns on again. Figure~\ref{Spin}b plots the signal strength as the "OFF" period (the recovery time) is varied, where the signal strength is the area under the transient curve of the fluorescence signal in Fig.~\ref{Spin}a, which is proportional to the number of unpolarized ions. We determine $\tau_{\rm spin}$ (the recovery time constant) by an exponential fit to the data. By measuring the linear dependence of $\gamma_{\rm spin}$ on $n_{\rm K}$, the total spin-relaxation rate coefficient $(\kappa_{\rm sd}+\kappa_{\rm se})$ can be acquired. For the $^{171}$Yb$^+$ ions, due to the optically resolved hyperfine splitting, the hyperfine optical pumping is easier to perform. The hyperfine polarization is achieved by pumping the ions into the $F=0$ state as shown in Fig.~\ref{Apparatus}c using $F=1$ to $F'=1$ transition. With some detailed calculations\cite{Happer2010}, we find that we can use the same signal recovery measurement procedure to obtain $(\kappa_{\rm sd}+\kappa_{\rm se})$. The experimental results are summarized in Fig.~\ref{Spin}c, which gives a mean value of $(\kappa_{\rm sd}+\kappa_{\rm se})=(3.1\pm0.92)\times10^{-9}$ cm$^3$s$^{-1}$ for both isotopes.

\begin{figure}[t]
	\includegraphics[trim=50 0 0 0,clip,scale=0.35]{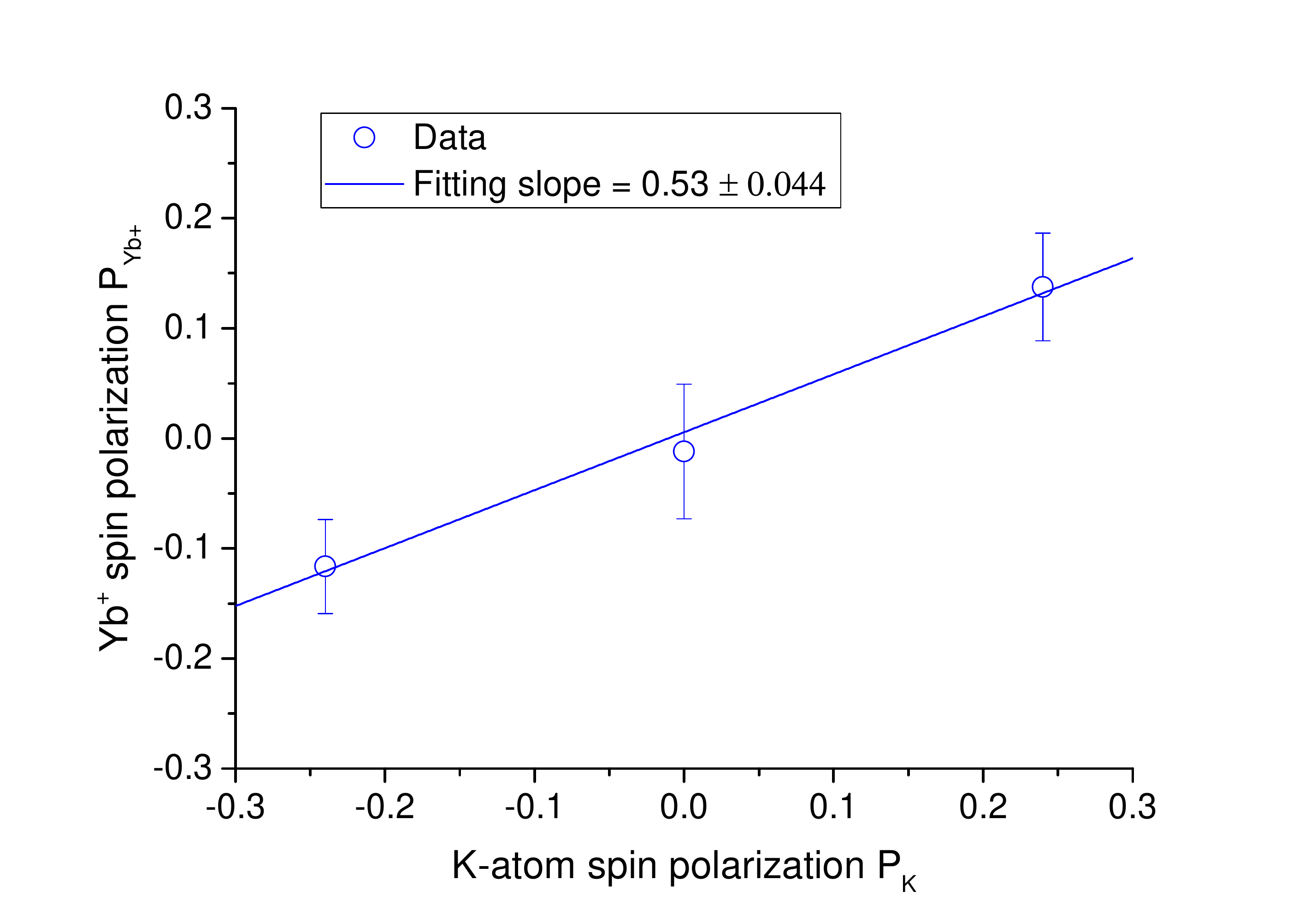}
\caption{Measured steady-state spin polarization of Yb$^+$ ions versus K-atom spin polarization. The fitting line is with 95\% confidence level, and its slope allows us to determine the spin-exchange rate coefficient.}
\label{PYbvsPK}
\end{figure}
Based on Eq.~\ref{SpinDynamics}, we can further determine $\kappa_{\rm sd}$ and $\kappa_{\rm se}$ separately by measuring the steady-state value of $P_{\rm Yb^+}$ during the ``OFF'' period through the following mathematical relation:
\begin{equation}\label{PYbPK}
P_{\rm Yb^+}=\frac{\kappa_{\rm se}P_{\rm K}}{\kappa_{\rm sd}+\kappa_{\rm se}}.
\end{equation}
The only difference in measurement settings is to have $P_{\rm K}\neq0$ and to have a long enough OFF period, which can be determined based on the $(\kappa_{\rm sd}+\kappa_{\rm se})$ value. To spin polarize the K atoms, we engaged a resonant 770-nm circularly polarized laser beam with a doughnut profile at the ion cloud, so the ions can be fit inside the dark volume (5-cm long, $>$ 2-mm diameter) of the doughnut profile to minimize the much larger charge-exchange effect between the Yb$^+$ ions and the excited K atoms when the 770-nm laser is present. The doughnut beam was generated using a single-lens imaging technique. The doughnut profile had a bright ring about 1-mm wide and 6.4 mW total optical power. The 770-nm laser linewidth was intentionally broadened to about 1-GHz to match up the Doppler-broadened K-atom $D1$ resonance. The measured, achievable maximum positive and negative K-atom spin polarization at the ion cloud inside the dark volume is $\pm0.24$ using the optical rotation method with a weak and optically detuned 766-nm beam. With these experimental conditions, Fig.~\ref{PYbvsPK} shows experimental data and a linear fit of $P_{\rm Yb^+}$ versus $P_{\rm K}$, where $P_{\rm Yb^+}$ can be well calibrated by measuring the signal levels of the ion fluorescence that are normalized to the case of unpolarized Yb$^+$ ions as illustrated in Fig.~\ref{Spin}b. We look at the normalized signal strength at the recovery time $\gg\tau_{\rm spin}$. We find that $P_{\rm Yb^+}=1$ when the Yb$^+$ signal magnitude is zero, $P_{\rm Yb^+}=0$ when the signal magnitude is one, and $P_{\rm Yb^+}=-1$ when the signal magnitude is two. The Yb$^+$ spin polarization $P_{\rm Yb^+}$ is a linear function of the fluorescence amplitude. Using Eq.~\ref{PYbPK} and the fitting slope from Fig.~\ref{PYbvsPK}, we find $\kappa_{\rm se}=(1.64\pm0.51)\times10^{-9}$ cm$^3$s$^{-1}$, and $\kappa_{\rm sd}\leq(1.46\pm0.77)\times10^{-9}$ cm$^3$s$^{-1}$.

Since we use spin-polarized K atoms to measure the spin-exchange rate, the measured $\kappa_{\rm se}$ can only be due to the interaction between Yb$^+$ ions and K atoms. With a relative collisional velocity $\bar{v}=4.6\times10^4$ cm/s determined by the Yb$^+$ ion and K atom temperatures, we find the effective spin-exchange cross section $\bar{\sigma}_{\rm se}=\kappa_{\rm se}/\bar{v}=(3.6\pm1.1)\times10^{-14}$ cm$^2$, which is similar to that of many alkali-alkali pairs \cite{Happer2010} in the thermal regime.  On the other hand,  $\kappa_{\rm sd}$ can have additional causes from other ion-gas-induced spin destruction if some gas background is still produced when the K-getter temperature increases (a similar concern for our charge-exchange experimental results). Therefore, we can only report the upper bound of $\kappa_{\rm sd}$ here.

\section{Discussion and Summary}
In the applications of using the atom-ion hybrid platform, we would like the spin-exchange rate to be much greater than the spin-destruction rate and the charge-exchange rate \cite{Major1968,tomza2018} to benefit quantum metrology, quantum information processing, and sympathetic cooling schemes. When comparing our experimental results with the Langevin rate coefficient $\kappa_{\rm L}=\pi\sqrt{4C_4/\mu}$, where $C_4$ is defined by the interatomic potential with a form of $-C_4/R^4$, and $\mu$ is the reduced mass, we find  $\kappa_{\rm L}=2.7\times10^{-9}$ cm$^3$s$^{-1}$ for a K-Yb$^+$ system. Hence we have  $\kappa_{\rm se}=(0.61\pm0.19)\kappa_{\rm L}$, $\kappa_{\rm ce}/\kappa_{\rm se}<10^{-4}$, and $\kappa_{\rm sd}/\kappa_{\rm se}\lesssim1$. The much smaller charge-exchange rate indicates that neutral K atoms can be used for sympathetic cooling of Yb$^+$ ions because of their favorable mass ratio and the relatively long Yb$^+$ ion trapping lifetime when K atoms are present. The spin-destruction rate is especially important in the quantum-physics applications, because it limits the coherence time of the internal states of the trapped ions with the presence of K atoms.

An earlier cold Rb-Yb$^+$ experiment \cite{Ratschbacher2013} demonstrated $\kappa_{\rm sd}/\kappa_{\rm se}\sim1.8$, which was able to be explained by a second-order spin-orbit (SO) model \cite{Tscherbul2016}. More recently, an experimental study on a cold Li-Yb$^+$ system\cite{Furst2018} demonstrates $\kappa_{\rm sd}/\kappa_{\rm se}\sim0.08$, and the work on a cold Rb-Sr$^+$ system \cite{Sikorsky2018} demonstrates $\kappa_{\rm sd}/\kappa_{\rm se}\sim0.19$. The much smaller $\kappa_{\rm sd}$ to $\kappa_{\rm se}$ ratio in the Li-Yb$^+$ system verifies the theoretical prediction of much weaker 2nd-order SO interactions with lighter neutral atoms \cite{Tscherbul2016,Kadlecek2001}. The theoretical work in Ref\cite{Tscherbul2016} also indicates a $T^{-0.3}$ temperature dependence of $\kappa_{\rm sd}$ for $T<0.3$ kelvin. If the same temperature dependence extends to $T\sim$ a few hundred kelvin, it should lead to a smaller $\kappa_{\rm sd}$ to $\kappa_{\rm se}$ ratio in the thermal regime. Unfortunately, our work can only set an upper limit of $\kappa_{\rm sd}$ and cannot provide a firm verification of the temperature dependence. But further detailed modeling of the spin-orbit interactions in the thermal regime will be helpful for understanding better the atom-ion collisional behaviors.

In conclusion, we have studied several important collisional effects in a thermal K-Yb$^+$ system. Our work may provide further insight to help develop atomic metrology, such as an ion clock without a UV light source, and study quantum physics using atom-ion hybrid platforms. Furthermore, K atoms can provide fermionic isotopes that may introduce additional advantages to the K$-$ion hybrid quantum systems.
\vspace{0pt} 
% use section* for acknowledgement
\section*{Acknowledgment}
We would like to thank Mr. Jeff Hunker for his help on setting up the experimental apparatus. This research was performed with funding from the Defense Advanced Research Projects Agency (DARPA). The views, opinions and/or findings expressed are those of the author and should not be interpreted as representing the official views or policies of the Department of Defense or the U.S. Government. Sandia National Laboratories is a multimission laboratory managed and operated by National Technology and Engineering Solutions of Sandia, LLC, a wholly owned subsidiary of Honeywell International, Inc., for the U.S. Department of Energy$'$s National Nuclear Security Administration under contract DE-NA0003525.

%\bibliographystyle{IEEEtran}
% argument is your BibTeX string definitions and bibliography database(s)
%\bibliography{IEEEabrv,../bib/paper}
%
% <OR> manually copy in the resultant .bbl file
% set second argument of \begin to the number of references
% (used to reserve space for the reference number labels box)
%\bibliographystyle{naturemag}
\bibliography{YbKrev}

% that's all folks
\end{document}